\documentclass[preprint,12pt]{elsarticle}

\usepackage{amssymb}
\usepackage{amsmath}
\usepackage{tabularx}

\journal{Journal of computation of physics}

\begin{document}

\begin{frontmatter}



\title{A difference-free conservative phase-field lattice Boltzmann method}


\author[a]{Chunheng Zhao}
\author[b]{Saumil Patel}
\author[c]{Taehun Lee}

\address[a]{Sorbonne Universit\'{e} and CNRS, Institut Jean Le Rond d'Alembert UMR 7190, F-75005 Paris, France}
\address[b]{Computational Science Division, Argonne National Laboratory}
\address[c]{Department of Mechanical Engineering, The City College of the City University of New York}

\begin{abstract}
We propose an innovative difference-free scheme that combines the one-fluid lattice Boltzmann method (lBM) with the conservative phase-field (CPF) lBM to effectively solve large-scale two-phase fluid flow problems. The difference-free scheme enables the derivation of the derivative of the order parameter and the normal vector through the moments of the particle distribution function (PDF). We further incorporate the surface tension force in a continuous surface stress form into the momentum equations by modifying the equilibrium PDF to eliminate the divergence operator. Consequently, the entire computation process, executed without any inter-grid finite difference formulation, demonstrates improved efficiency, making it an ideal choice for high-performance computing applications. We conduct simulations of a single static droplet to evaluate the intensity of spurious currents and assess the accuracy of the scheme. We then introduce the density or viscosity ratio and apply an external body force to model the Rayleigh-Taylor instability and the behavior of a single rising bubble, respectively. Finally, we employ our method to study the phenomenon of a single bubble breaking up in a Taylor-Green vortex. The comparison between the difference-free scheme and the finite difference method demonstrates the scheme's capability to yield accurate results. Furthermore, based on the performance evaluation, the current scheme exhibits an impressive $47\% $ increase in efficiency compared to the previous method.
\end{abstract}



\begin{highlights}
\item We propose a new LB scheme to solve the two-phase fluid flow system without derivative calculation.
\item The mass-conserving character of the new scheme is validated.
\item The new scheme highly improves the efficiency for large-scale emulsion problems.
\end{highlights}


\end{frontmatter}

\section{Introduction}

Emulsions, which involve the multi-phase flows of two immiscible fluids with similar densities~\cite{crialesi2022modulation}, have found numerous applications in industry such as froth flotation~\cite{fuerstenau2007froth,rao2013surface}, medicine delivery~\cite{panigrahi2021quality,rajian2011drug,aditya2015co}, and oil production~\cite{kokal2005crude,mandal2010characterization,kilpatrick2012water}. The behaviors of emulsions, such as intricate surface deformation, droplet breakup, and coalescence, pose challenges to numerical simulations. The methodology for modeling the surface effect in this case is of significant importance. When simulating the surface force using the continuum surface force (CSF) method~\cite{brackbill1992continuum}, the presence of small droplets with large curvature can lead to computational instability. The continuous surface stress (CSS) method improves the numerical stability since the curvature evaluation is not essential compared to the CSF~\cite{lafaurie1994modelling}. However, CSS suffers from the problem of spurious currents~\cite{lafaurie1994modelling}. Based on the diffuse interface method, potential form surface force formulation can alleviate the difficulty of the curvature calculation as well, and it is able to eliminate the spurious currents to round off~\cite{jacqmin1996energy,kim2005continuous,lee2006eliminating}. However, in this case, mass loss is a significant concern due to the large curvature effect~\cite{crialesi2022modulation,zheng2014shrinkage,yue2007spontaneous,perlekar2012droplet,biferale2011lattice}. Another difficulty of emulsion simulation is that conducting large-scale three-dimensional simulations requires extensive computational resources and time. Normally, when the derivative computation is needed, one has to employ the message-passing interface (MPI) library~\cite{gropp1999using} to transfer the essential data between multiple computers or memory which is time-consuming. 

Numerous numerical studies have been conducted on emulsions using the diffuse interface method, but a recurring issue encountered is the problem of mass loss~\cite{crialesi2022modulation}. Using the pseudopotential lattice Boltzmann method (lBM), droplet statistics and correlation with turbulence were systematically investigated~\cite{perlekar2012droplet,haakansson2022deformation}. While in those studies, significant mass loss was found due to the droplet dissolution. The Cahn-Hilliard (C-H) method was employed for modeling emulsions of binary flow~\cite{perlekar2014spinodal}. In this method, a finite interface thickness is considered, and the evolution of the interface is captured by a fourth-order partial differential equation. Although the C-H equation is in a so-called conservative form, it suffers from the mass loss problem for the simulation of small droplets~\cite{yue2007spontaneous,zheng2014shrinkage}. Currently, the conservative phase-filed (CPF) method is utilized to solve the interface evolution using a second-order partial differential equation. Unlike the C-H method, the CPF method eliminates the curvature-driven term, enhancing its conservative nature for simulations involving large curvature morphologies~\cite{chiu2011conservative}. This characteristic has been particularly beneficial in studies that focus on the mass loss of large curvature droplets by using the C-H method~\cite{yue2007spontaneous,zheng2014shrinkage,geier2015conservative,baroudi2021simulation,zhao2023interaction}. Additionally, the CPF method achieves higher efficiency compared to the C-H method, as it avoids higher-order derivative calculations. Geier et al.~\cite{geier2015conservative} utilized lBM to solve CPF. Based on this method, ternary fluid flow and adaptive mesh refinement technology have also been applied to the lBM CPF~\cite{zhao2022ternary,zhao2023engulfment,zhao2023interaction, fakhari2016mass}. Using asymptotic analysis on a diffusive scale~\cite{junk2005asymptotic}, the CPF lBM can evaluate the derivative of the order parameter $\phi$ and the normal vector $\mathbf{n}$ from the central moment of the PDF~\cite{geier2015conservative,geier2006cascaded,geier2015cumulant}, up to $\mathcal{O} (\epsilon^2)$, where $\epsilon\sim\Delta x $ or $\epsilon^2\sim \Delta t$ in diffusive scale.

The lBM has been widely validated as a solver for multi-phase flow simulations using different models. In the single-fluid model, it has been successfully employed to solve the Navier-Stokes (N-S) equations, where phases are separated by pressure. Alternatively, the multi-fluid model couples the N-S equations with phase-field equations, enabling an accurate representation of phase separation~\cite{lee2006eliminating,lee2008wall,shan1993lattice,he1999lattice,lee2005stable,lee2010lattice,geier2015conservative}. The evolution of the particle distribution function (PDF) in the lBM follows a two-step process: propagation and collision. This localized computation of the collision operator makes the lBM well suited for parallel computing. Additionally, the advection, including the material derivative of the PDF, is typically solved using the Crank-Nicholson method, a semi-implicit finite difference approach that ensures numerical stability. An advantageous feature of the lBM is its ability to implicitly recover derivatives without the need for additional finite difference calculations~\cite{lallemand2000theory,he1997lattice,dellar2002lattice,reis2022lattice}. For instance, the original ideal gas lBM computes the pressure gradient and viscous stress force using the moments of the PDF. During simulation, the propagation step provides crucial information for derivative calculations. By applying specific assumptions and considering certain limits, we are able to reconstruct the equilibrium PDF and extract valuable information such as the surface stress, density derivative, and normal vector \cite{reis2022lattice,geier2015conservative}.

A recent advancement by Reis introduced a one-fluid model that addresses the implicit calculation of surface stress by modifying the equilibrium PDF \cite{reis2022lattice}. By neglecting the high-order error ($\mathcal {O}(Ma^2/Re)$), the N-S momentum equation with the CSS formulates the surface force, allowing for the recovery of pressure evolution. Since the divergence of the surface stress is implicitly solved, the efficiency is highly improved. We here propose to combine the one-fluid lBM model with the CPF lBM \cite{geier2015conservative} to construct a difference-free lBM to solve the binary fluid flow. In this method, unlike the advection-diffusion-sharpening lBM~\cite{reis2018conservative}, the normal vector and density gradient are implicitly recovered using the central moment, which can be obtained through the CPF lBM. By combining these two methods, we can effectively solve binary fluid flow without the need for the finite difference method. All derivative computations are implicitly resolved using the moments of the PDF.

The difference-free method proposed in this study is rigorously validated through a series of tests. In the single-droplet simulation, we assess the accuracy and reliability of the method, and it demonstrates excellent agreement with established research in this field. Subsequently, we expand the scope by incorporating density ratio and viscosity ratio parameters, allowing us to successfully simulate both the Rayleigh-Taylor instability and the single rising bubble benchmarks. These test cases further confirm the effectiveness and consistency of our method, aligning well with previous findings. Furthermore, we investigate the challenging scenario of a single bubble droplet breakup within a Taylor-Green vortex cube using the derived difference-free scheme. Using the localized evaluation of derivatives, our simulation exhibits enhanced efficiency, especially for large-scale simulations. In particular, the results obtained from this problem show strong consistency with our previously developed approach, providing additional confidence in the accuracy and robustness of the difference-free method.

\section{Numerical Methodology}
\subsection{governing equations}
The governing equations consist of the Navier-Stokes equations augmented with the continuous surface stress tensor, which represents the surface tension stress, and the conservative phase-field equation. Mathematically, these equations can be expressed as follows:
\begin{equation}\label{ns_c}
    \frac{1}{\rho c_s^2}\frac{\partial p}{\partial t}+\nabla\cdot\mathbf{u}=0,
\end{equation}
\begin{equation}\label{ns_m}
    \frac{\partial\rho\mathbf{u}}{\partial t}+\nabla\cdot(\rho\mathbf{u}\otimes\mathbf{u})=
    \nabla\cdot
    \left(\mathbf{\Pi_{v}}+\mathbf{\Pi_{s}}-p\mathbf{I}
    \right)+\rho\mathbf{G},
\end{equation}
\begin{equation}\label{cac}
    \frac{\partial\phi}{\partial t}+\nabla\cdot(\phi\mathbf{u})=
    \nabla\cdot
    M\left(\nabla\phi-\frac{4\phi(1-\phi)}{\delta }\mathbf{n}
    \right).
\end{equation}
In the given context, we utilize several symbols and terms to represent different physical quantities. Eq.~(\ref{ns_c}) represents the pressure evolution equation, where $\rho$ denotes the local density, $p$ represents the pressure, $\mathbf{u}$ signifies the external velocity vector and $c_s$ corresponds to the speed of sound. In the low Mach number regime ($Ma=|\mathbf{u}|/c_s\ll1$), it is valid to assume that the velocity field is approximately divergence-free, as established in previous studies \cite{lee2005stable,reis2022lattice}. On the other hand, Equation (\ref{ns_m}) introduces the tensor product operation denoted by $\otimes$, the identity tensor represented by $\mathbf{I}$, and the external body forcing term denoted by $\rho\mathbf{G}$. The terms $\mathbf{\Pi_v}$ and $\mathbf{\Pi_s}$ correspond to the viscous stress and the continuous surface stress, respectively, and they can be expressed as follows:
\begin{equation}
    \mathbf{\Pi_{v}}=     \eta\left(\nabla \mathbf{u}+(\nabla\mathbf{u})^T\right),
\end{equation}

\begin{equation}\label{surface}
    \mathbf{\Pi_{s}}= \sigma|\nabla\phi|\left(\mathbf{I}-\mathbf{n} \otimes\mathbf{n}\right).
\end{equation}
Here, $\eta$ denotes the local viscosity, $\delta$ represents the interface thickness, $\sigma$ corresponds to the surface tension between the two phases, and $\mathbf{n}$ represents the normal vector. In Equation (\ref{cac}), the order parameter $\phi$ takes on values within the range of $[0,1]$, serving as a distinguishing parameter between different fluids. Specifically, within the interface region, $0<\phi<1$, the order parameter varies, while in the bulk fluids, it takes either the value $\phi=0$ or $\phi=1$. Additionally, in Equation (\ref{cac}), the coefficient $M$ is referred to as the diffusion coefficient or mobility.

Eqs. (\ref{ns_m}) and (\ref{cac}) in their current form, without the inclusion of a body force, are formulated in what is commonly referred to as the conservative form. This particular formulation is advantageous as it ensures the preservation of both momentum and mass throughout the simulation \cite{zhao2023interaction}. Using a Chapman-Enskog analysis with convective scaling, it becomes possible to recover the stress force, which includes the contribution of surface tension \cite{he1997lattice,dubois2008equivalent,reis2022lattice,junk2005asymptotic}. A comprehensive derivation of the conservative phase-field equation given by Eq. (\ref{cac}), can be found in previous work such as \cite{geier2015conservative,zhao2023interaction}. Moreover, a comparative analysis of the mass conservation characteristics between the Cahn-Hilliard method and the conservative phase-field method has been explored in previous research \cite{baroudi2021simulation}.

\subsection{one-fluid lattice Boltzmann model for N-S equations}
Governing Eqs. (\ref{ns_c}) and (\ref{ns_m}) are solved using the recently developed one-fluid lattice Boltzmann method (lBM) as proposed by Reis \cite{reis2022lattice}. In this approach, the surface tension stress is incorporated into the equilibrium particle distribution function (PDF). Notably, the divergence term in Equation (\ref{ns_m}) associated with surface stress is implicitly resolved by utilizing moments of the PDF.

We start with the discrete Boltzmann equation of the PDF $f_i$ with an external forcing term:
\begin{equation}\label{dbe_m}
    \left(\frac{\partial}{\partial t}+\mathbf{e}_i\cdot\nabla\right)f_i=-\frac{1}{\lambda_\rho}(f_i-f_i^{eq})+F_i.
\end{equation}
The PDF restriction is given as $\sum_i f_i=p/c_s^2$. In the above equation, $\mathbf{e}_i$ denotes the discrete velocity vector in a $D2Q9$ lattice~\cite{lee2005stable}, which is given as:
$$ \boldsymbol{e}_i=\left\{
\begin{aligned}
&(0,0),&     &i=0 \\
&(\cos\theta_i,\sin\theta_i),& \theta_i=(i-1)\pi/2, \quad &i=1,2,3,4\\
&\sqrt{2}(\cos\theta_i,\sin\theta_i),& \theta_i=(i-5)\pi/2+\pi/4, \quad &i=5,6,7,8,
\end{aligned}
\right.
$$
The parameter $\lambda_\rho$ represents the relaxation time, which is connected to the local kinematic viscosity $\nu$ through the equation $\nu=\lambda_\rho c_s^2$, where $c_s=1/\sqrt{3}$. The external forcing term $F_i$ is comprised of two components: the compensation part $S_i$ (provided in \ref{ap1}), and the external body force part:
\begin{equation}\label{body}
    R_i=t_i(\frac{\mathbf{e}_i-\mathbf{u}}{c_s^2}+\frac{\mathbf{e}_i\cdot\mathbf{u}}{c_s^4}\mathbf{e}_i)\cdot(\rho\mathbf{G}).  
\end{equation}

As demonstrated in the work by Reis \cite{reis2022lattice}, the equilibrium PDF $f^{eq}_i$, which incorporates the surface stress, can be expressed as follows:
\begin{equation}\label{geq_s}
    f_i^{eq}=t_i\left(\frac{p}{c_s^2}+\rho\left( \frac{\mathbf{e}_i \cdot \mathbf{u}}{c_s^2}+
\frac{(\mathbf{e}_i\cdot\mathbf{u})^2}{2c_s^4}-
\frac{|\mathbf{u}|^2}{2c_s^2}\right) +\frac{1}{2c_s^4}\mathbf{\Pi}_s \mathbf{:} \left(\mathbf{e}_i\otimes\mathbf{e}_i-c_s^2\mathbf{I}\right)\right).
\end{equation}
Here, the weights $t_i$ associated with the equilibrium PDF have specific values: $t_0=4/9$, $t_1=t_3=t_5=t_7=1/9$, and $t_2=t_4=t_6=t_8=1/36$. Notably, through the modification of the equilibrium PDF, it is observed that the explicit divergence operator is implicitly resolved up to $\mathcal{O}(Ma^2)$ accuracy. Further details and a comprehensive Chapmann-Enskog analysis of this aspect can be found in the appendix.

Using the Crank-Nicolson finite difference method, the discrete Boltzmann equation given by Eq. (\ref{dbe_m}) can be effectively solved through the application of the Lattice Boltzmann method~\cite{zhao2023interaction}:
\begin{equation}\label{c-n_m}
    \bar{f}_i(\mathbf{x}+\Delta t \mathbf{e}_i,t+\Delta t)-\bar{f}_i(\mathbf{x},t)=-\frac{1}{\tau_\rho+0.5}(\bar{f}_i(\mathbf{x},t)-\bar{f}_i^{eq}(\mathbf{x},t))+\Delta t F_i,
\end{equation}
where $\tau_\rho=\lambda_\rho/\Delta t$ denotes the dimensionless relaxation time, and the modified equilibrium PDF is given as $\bar{f}_i^{eq}=f_i^{eq}-0.5\Delta t F_i$.

The solution to Eq.~(\ref{c-n_m}) is obtained by performing a local operator collision and a non-local operator propagation. The after-collision PDF, denoted as $f_i^*$, can be calculated using the following equation:
\begin{equation}
    f_i^*=\bar{f}_i-\frac{1}{\tau_\rho+0.5}(\bar{f}_i-\bar{f}_i^{eq})+\Delta t F_i.
\end{equation}
It is followed by a Lax–Wendroff propagation method:
\begin{flalign*}
    \bar{f}_i(\mathbf{x},t+\Delta t)=f_i^*(\mathbf{x},t)-c\left[f_i^*(\mathbf{x},t)-f_i^*(\mathbf{x}-  \mathbf{e}_i,t)\right]\\-0.5c(1-c)\left[f_i^*(\mathbf{x}+\Delta t\mathbf{e}_i,t)-2f_i^*(\mathbf{x},t)+f_i^*(\mathbf{x}-\Delta t\mathbf{e}_i,t)\right],
\end{flalign*}

\begin{flalign*}
    \bar{f}_i(\mathbf{x}+\Delta t\mathbf{e}_i,t+\Delta t)=    \bar{f}_i(\mathbf{x},t+\Delta t),
\end{flalign*}
where the Courant number $ c=0.997$ is fixed in our current work.  
Finally, the pressure and momentum can be recovered by the PDFs by: 
\begin{equation}\label{pressure}
    p=c_s^2\sum_i \bar{f}_i+\frac{c_s^2\Delta t}{2} \mathbf{u}\cdot\nabla\rho,
\end{equation}
\begin{equation}\label{momentum}
    \rho\mathbf{u}=\sum_i \bar{f}_i \mathbf{e}_i-\frac{\Delta t}{2} \rho\mathbf{G}.
\end{equation}
\subsection{conservative phase-field lattice Boltzmann method}
The conservative phase-field equation is solved using the lattice Boltzmann scheme described in Geier et al. \cite{geier2015conservative}. The discrete Boltzmann equation for the PDF of the order parameter $g_i$ can be expressed as follows:
\begin{equation}\label{dbe_order}
    \left(\frac{\partial}{\partial t}+\mathbf{e}_i\cdot\nabla\right)g_i=-\frac{1}{\lambda_\phi}(g_i-g_i^{eq}),
\end{equation}
with
\begin{equation}\label{geq}
    g_i^{eq}=t_i\phi\left[1+\frac{\mathbf{e}_i\cdot\mathbf{u}}{c_s^2}+\frac{(\mathbf{e}_i\cdot\mathbf{u})^2}{2c_s^4}-\frac{|\mathbf{u}|^2}{2c_s^2}\right]+t_i\mathbf{e}_i\cdot\mathbf{S}. 
\end{equation}
The source term $\mathbf{S}$ is incorporated to account for the separation flux term in Eq.~(\ref{cac}). It can be expressed as follows:
\begin{equation}
    \mathbf{S}=\frac{4\phi(1-\phi)}{\delta}\mathbf{n}.
\end{equation}
By means of the Crank-Nicholson method, the lattice Boltzmann equation can be shown as:
\begin{equation}\label{c-n_c}
    g_i(\mathbf{x}+\Delta t \mathbf{e}_i,t+\Delta t)-g_i(\mathbf{x},t)=-\frac{1}{\tau_\phi+0.5}\left(g_i(\mathbf{x},t)-g_i^{eq}(\mathbf{x},t)\right),
\end{equation}
where $\tau_\phi=\lambda_\phi/\Delta t$ denotes the relaxation time which is related to mobility $M=\tau_\phi c_s^2\Delta t$. Similar to the one-fluid lBM, the lattice Boltzmann equation given by Eq.~(\ref{c-n_c}) is solved using a collision and Lax-Wendroff propagation method, with a chosen Courant number $c=0.997$. The order parameter $\phi$ can be computed as the summation of $g_i$ values, while the local density $\rho$ is determined as $\rho=\rho_1\phi+\rho_2(1-\phi)$.

The density gradient $\nabla\rho$ and the normal vector $\mathbf{n}$ in Eq.~(\ref{surface}) and (\ref{pressure}) can be determined using the central moment of the PDF, as demonstrated in Geier et al. \cite{geier2015conservative}. It has been established that the normal vector and the derivative of the order parameter can be accurately recovered through the central moment approximation up to $\mathcal{O}(Ma^3)$. Within this method, we represent the first-order central moment vector as $\mathbf{K}_1$, which can be expressed as follows:
\begin{equation}
    \mathbf{K}_1=\sum_i g_i(\mathbf{e}_{i}-\mathbf{u} ).
\end{equation}
The normal vector can be recovered from the central moment thanks to the lattice Boltzmann method:
\begin{equation}\label{norm}
    \mathbf{n}=-\frac{\mathbf{K}_1}{|\mathbf{K}_1|}.
\end{equation}
Furthermore, once we have the expression for the normal vector, we can obtain the first derivative of the order parameter as follows:
\begin{equation}
    \nabla\phi=\frac{1}{(\tau_\phi+0.5)c_s^2}\left(\tau_\phi c_s^2\mathbf{S}-\mathbf{K}_1\right).
\end{equation}
The density gradient can be calculated as $\nabla\rho=\Delta\rho\nabla\phi$, where $\Delta \rho=\rho_1-\rho_2$ represents the density difference between the two fluids. The detailed derivation of this expression is provided in Appendix \ref{ap2}.

Finally, let us summarize the computation process of this difference-free method. The order parameter $\phi$ is calculated from the moments of the PDF $g_i$, while the gradient $\nabla\phi$ and the normal vector $\mathbf{n}$ are obtained from the central moment $\mathbf{K}_1$. With these values, we can then recover the pressure and the momentum of the fluid flow using Eq.~(\ref{pressure}) and (\ref{momentum}).

\section{Simulation results}

\subsection{Single static droplet}
The objective of the single static droplet simulation is to assess the accuracy and stability of the difference-free scheme. As the initial droplet morphology deviates from a perfect circle, the interface undergoes deformation driven by curvature differences until it reaches equilibrium. It is worth noting that inconsistent discretization schemes or imbalanced surface tension force formulations can lead to the emergence of spurious currents, as described in previous studies~\cite{lee2006eliminating, popinet2018numerical}.

\begin{table}
\caption{\label{Tab:1}Convergence test of the spurious currents' intensity with  different $La$.}
\begin{tabular}{llllll}
   &
  \multicolumn{2}{l}{$\boldsymbol{u}_{max}^2$}&\\
 \cline{2-5}
 $La$ &CSF& Potential form & FDCSS & difference-free scheme\\ \hline

  16    & $8.5\times10^{-14}$ &$1.1\times10^{-12}$ & $3.8\times10^{-12}$ & $2.3\times 10^{-11}$\\

  4     & $2.2\times10^{-14}$ &$3.2\times10^{-13}$ & $1.1\times10^{-12}$ & $1.5\times 10^{-11}$\\

  1     & $5.3\times10^{-15}$ &$1.1\times10^{-13}$ &$4.3\times10^{-13}$ & $8.8 \times 10^{-12}$ \\

  0.25  & $      2.4\times10^{-15}$ &$5.1\times10^{-14}$ &$1.8\times10^{-13}$& $6.5\times 10^{-12}$\\

\end{tabular}
\end{table}
\begin{figure}
	\centering
  \includegraphics[width=\linewidth]{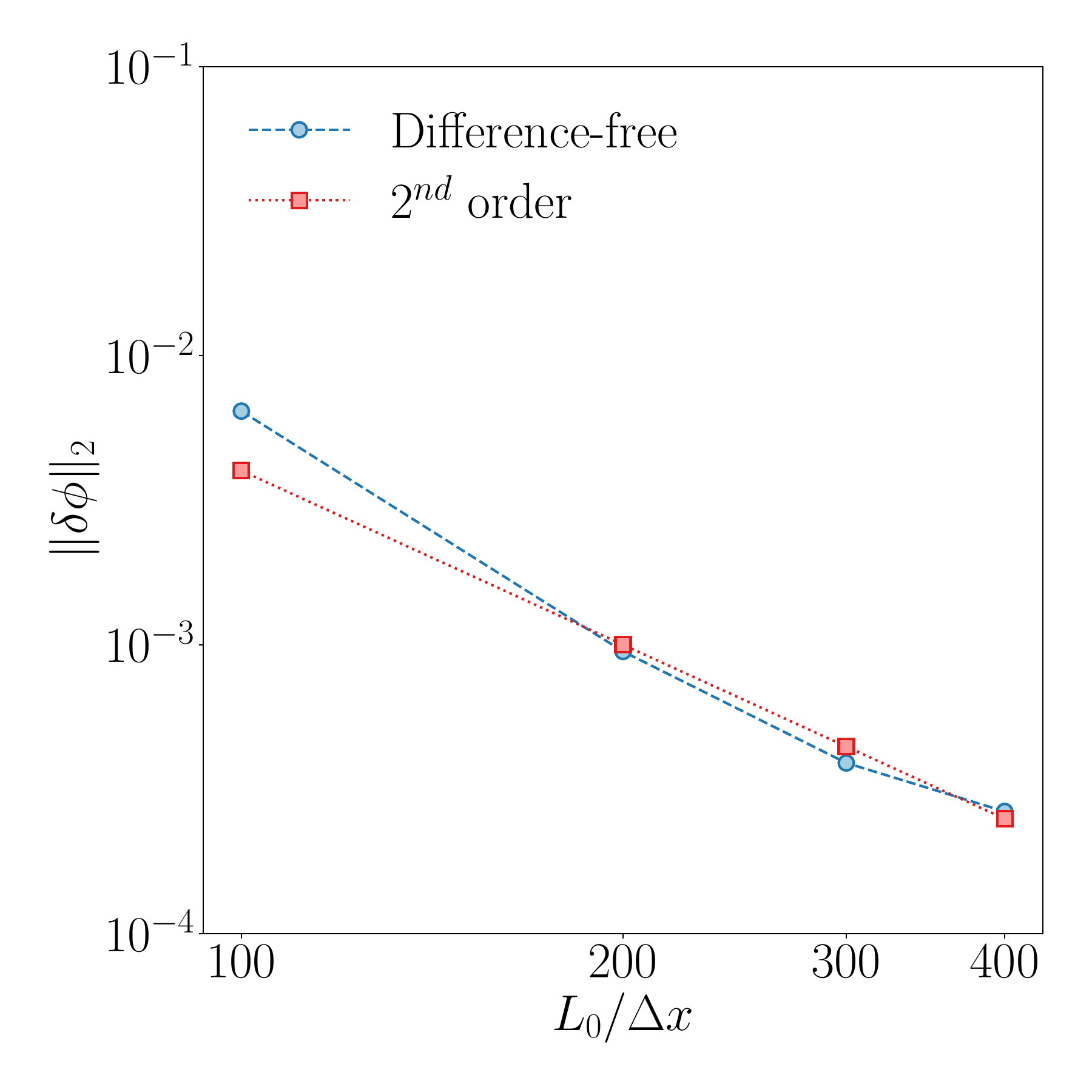}
    \caption{\label{converge} Convergence trend of the $\Vert \delta\phi\Vert_2$ for single droplet simulation with $Cn=0.06$, $La=1$ and $L_0/\Delta x=[100,400]$. The blue dashed line indicates the $2^{nd}$ order reference line.
    }
\end{figure}

In the 2-D or 3-D droplet simulation, Eq.(\ref{cac}) provides an approximate solution rather than an exact one-dimensional solution. Consequently, after initialization, the droplet undergoes further deformation due to this approximation. The interface deformations induce spurious currents, which are eventually dissipated by viscous effects, leading to a pressure equilibrium. In our previous work, we evaluated three surface force formulations for the single static droplet: continuum surface force (CSF)\cite{brackbill1992continuum, kim2005continuous}, finite difference based continuous surface stress (FDCSS)~\cite{lafaurie1994modelling}, and potential form surface force~\cite{jacqmin1996energy, junk2005asymptotic}. In this study, we apply the same parameters and compare the results with the previous work using the difference-free scheme. The Laplace number, $La = \sigma \rho D/\eta$, is used to characterize the balance between surface effects and viscous effects. In this test, a droplet with a diameter $D$ is initialized at the center of a square domain with a length of $L_0 = 2D$. The density ratio and viscosity ratio are fixed at $\rho_1/\rho_2=1$ and $\eta_1/\eta_2=1$. It is important to note that in this comparison, the only difference between the current numerical scheme and the previous method described in \cite{zhao2023interaction} lies in the surface effect. The droplet shape is described by the hyperbolic tangent function:
\begin{equation}\label{initial_droplet}
    \phi=\frac{1}{2}\left[1+\tanh{\frac{2(|\mathbf{z}|-D/2)}{\delta}}\right].
\end{equation}
The interface thickness $\delta$ is characterized by the Cahn number $Cn = \delta/D = 0.06$. The distance from a point to the droplet interface is denoted by $|\mathbf{z}|-D/2$, where $|\mathbf{z}|$ represents the distance between the local point axis and the center of the droplet. In this test, we apply periodic boundary conditions. The comparison of the four formulations with different $La$ values is presented in Table~\ref{Tab:1}. The intensity of the spurious currents is computed by $|\mathbf{u}_{max}|^2$, where $|\mathbf{u}_{max}|$ denotes the maximum norm of the velocity vector. Those values are obtained after $200T/t_0$ to ensure the simulation reaches a steady state, and  $t_0=\eta D/2\sigma$ is the viscous time scale~\cite{zhao2023interaction}. As discussed in \cite{zhao2023interaction}, the CSF formulation compensates for the curvature-driven effect, which is subtracted in the conservative phase-field equation. As a result, the intensity of spurious currents is smaller in this formulation compared to the other three formulations or schemes. The difference-free scheme yields slightly worse results than the previous approach; however, it exhibits lower spurious currents' intensity as the $La$ value decreases. It has been argued that the difference-free scheme requires a relatively thicker interface to fully resolve the interface dynamics \cite{geier2015conservative}. Therefore, when dealing with interfaces undergoing large deformations, we suggest using a higher resolution to ensure accurate simulations.

To further assess the accuracy of the difference-free scheme, we perform a convergence study by comparing the relative error between the numerical results and the analytic solution. The relative error is calculated using the following formula:
\begin{equation}
    ||\delta \phi||_2=\sqrt{\frac{\sum_{x,y}(\phi-\phi_0)^2}{\sum_{x,y}\phi_0^2}}.
\end{equation}

Here, we set the initial order parameter profile of the droplet as $\phi_0$. We maintain a fixed relaxation time $\tau_\rho=0.02$ and Laplace number $La=1$, while varying the resolution $L_0/\Delta x$ in the range of $100$ to $400$. The interface thickness in the test remains fixed at $Cn=0.06$. The convergence study results are depicted in Figure~\ref{converge}, where it can be observed that as the resolution increases while keeping the same $Cn$, the relative error decreases and exhibits second-order accuracy. Notably, for interface thicknesses $\delta/\Delta x > 4$, a better-resolved simulation is expected.

\subsection{Rayleigh-Taylor instability}
The Rayleigh-Taylor instability is a well-known benchmark problem extensively studied in the field of two-phase flow~\cite{he1999lattice,reis2022lattice,PhysRevE.87.043301}. In contrast to the single droplet simulation, this test introduces an external forcing term in the momentum equations, as shown in Eq.~(\ref{body}), while also considering the effect of density ratio. The objective of this test is to analyze the behavior of the interface between two fluids subjected to gravitational acceleration and evaluate the performance of the difference-free scheme in capturing the instabilities and interface dynamics.
\begin{figure}
	\centering
  \includegraphics[width=\linewidth]{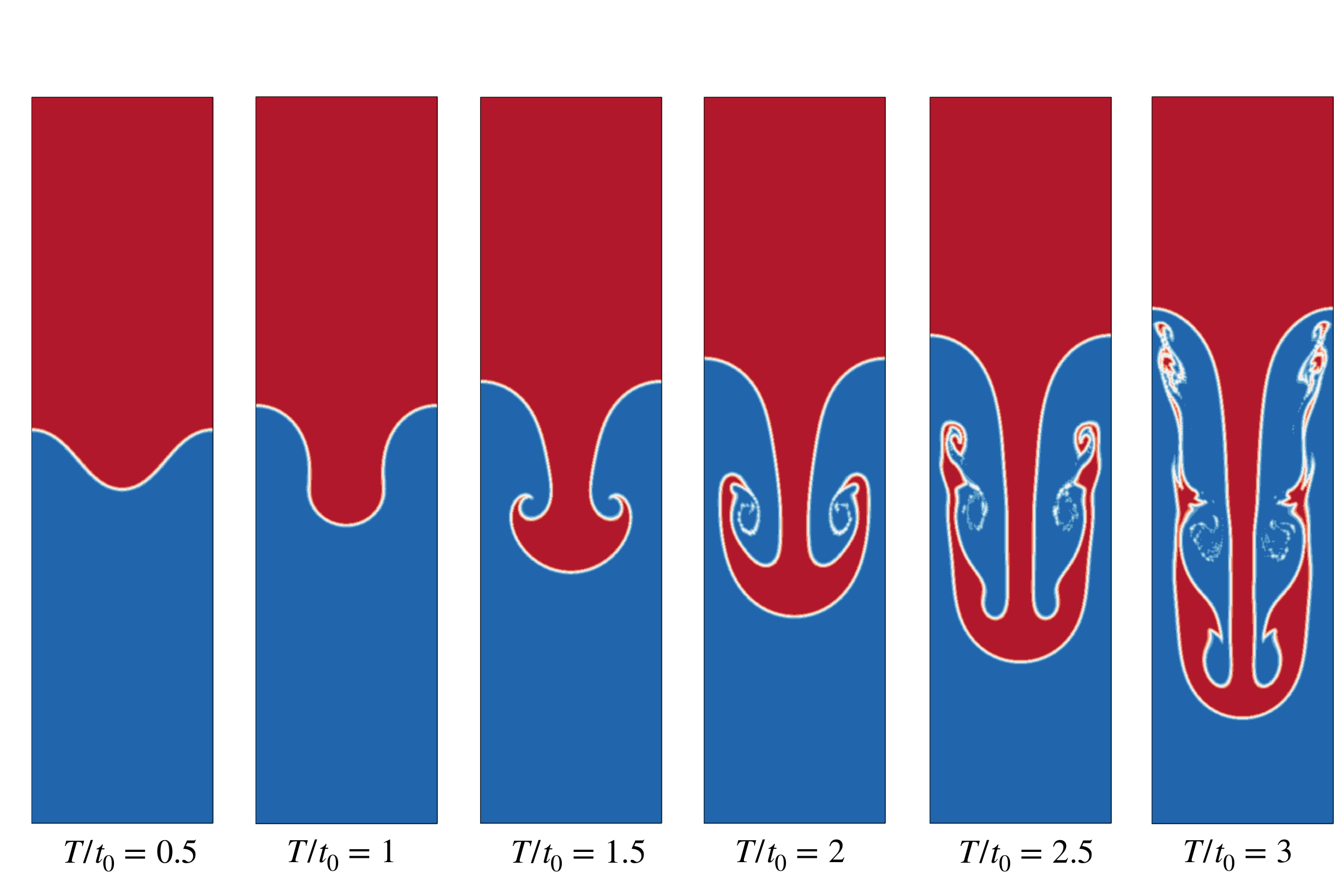}
    \caption{\label{RTI} Contour of the order parameter for Rayleigh-Taylor instability at time $T/t_0=[0.5,3]$ by using the derivative free scheme. In this case, $Re=2000$, $Ca=0.1$, $\rho_1/\rho_2=3$, $\eta_2/\eta_1=3$. 
    }
\end{figure}

\begin{figure}
	\centering
  \includegraphics[width=\linewidth]{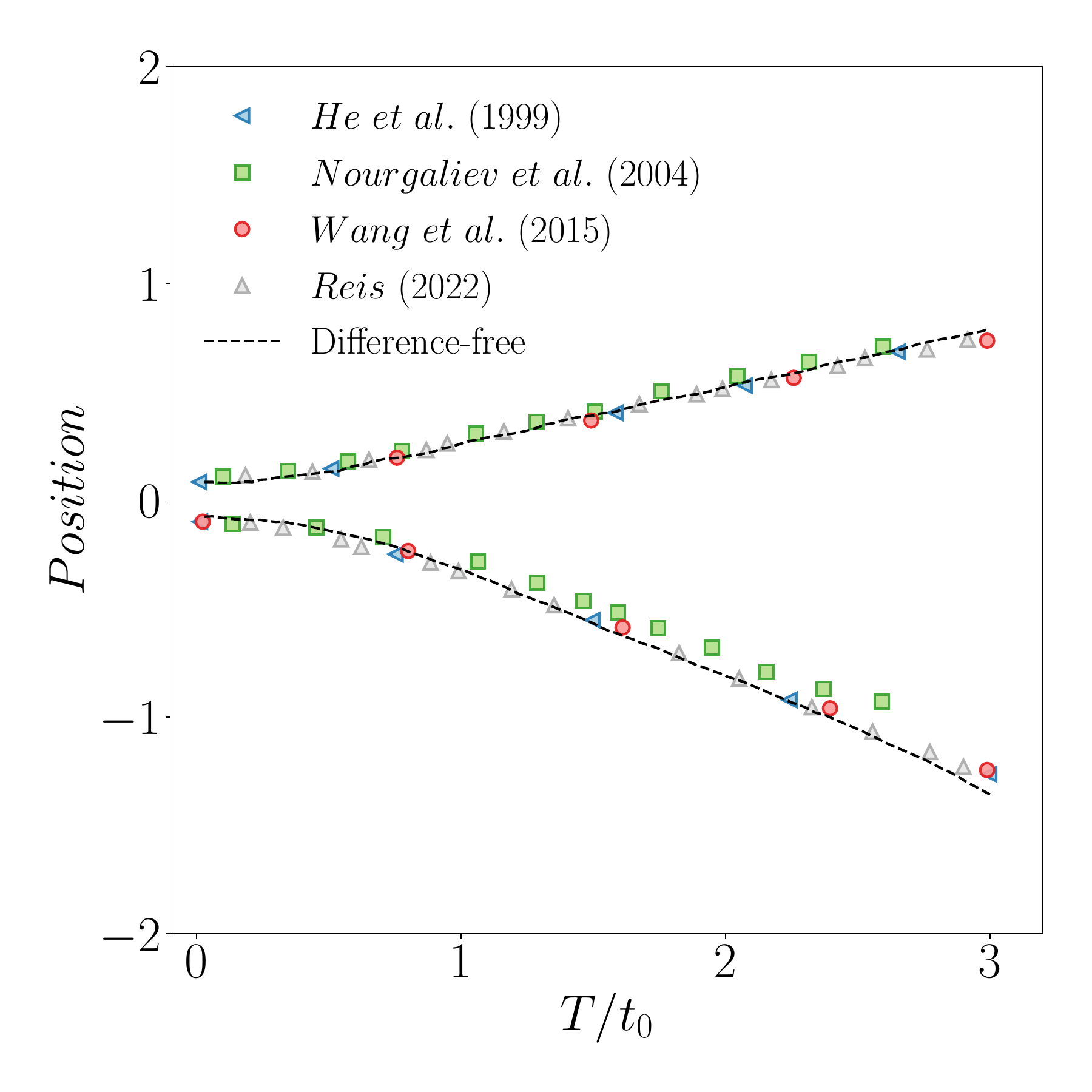}
    \caption{\label{rt} Comparison of the evolution of the peak interface where $\phi=0.5$ at $x/L_0=0$ (top line), and trough interface $\phi=0.5$ at $x/L_0=0.5$ (bottom line) positions between numerical results with previously published research.
    }
\end{figure}
In the Rayleigh-Taylor instability test, we initialize two fluids in a rectangular pool with a domain length of $L_0\times 4L_0$. The initial position of the interface is determined by the equation $y/L_0=0.1 \cos\left(2\pi x/L_0\right)$, and two center lines of the rectangular pool are indicated by $x/L_0=0.5$ and $y/L_0=0$. We set the density ratio and viscosity ratio between the top and bottom fluids as $\rho_1/\rho_2=3$ and $\eta_1/\eta_2=1$, respectively, which allows us to calculate the Atwood number as $At=(\rho_1-\rho_2)/(\rho_1+\rho_2)=0.5$. To characterize the simulation, we introduce the Reynolds number $Re=\rho_1 U L_0/\eta_1=2000$, where $U=\sqrt{gL_0}$ represents the velocity scale. Additionally, we consider the surface effect by incorporating the capillary number $Ca=\eta_1 U/\sigma=0.1$. These parameters enable us to assess the behavior and evolution of the interface under the influence of gravitational acceleration and surface tension.

In Figure~\ref{RTI}, we present the evolution of the Rayleigh-Taylor instability for a time range of $T/t_0=[0.5,3]$, where $t_0=\sqrt{2L_0/g}$. As anticipated, the instability progresses over time, resulting in complex interfacial dynamics. To assess the accuracy of our method, we compare the evolution of the interface position with previous studies, as illustrated in Figure~\ref{rt}. The agreement between our results and those from other methods is excellent, both for the peak and trough positions of the interface. This demonstrates the reliability and effectiveness of our approach in capturing the essential features of the Rayleigh-Taylor instability.

\subsection{Rising Bubble}

\begin{figure}
	\centering
  \includegraphics[width=\linewidth]{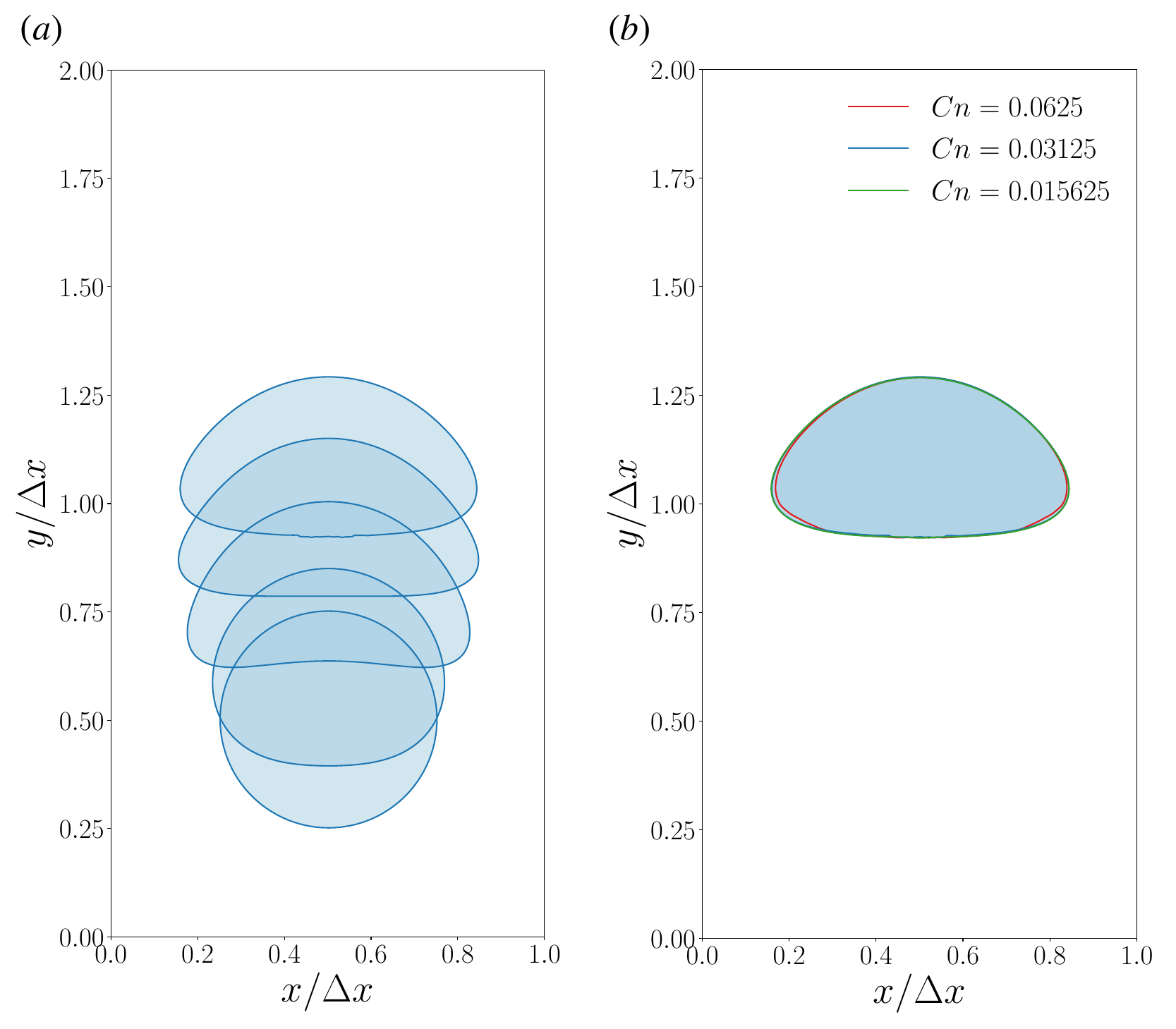}
    \caption{(a) Evolution of the rising bubble in equal-sized time steps. (b) Comparison of the rising bubble morphology at $T/t_0=3$ for $Cn=[0.0625,0.015625]$. \label{rising}  \ 
    }
\end{figure}
\begin{figure*}
	\centering
  \includegraphics[width=\linewidth]{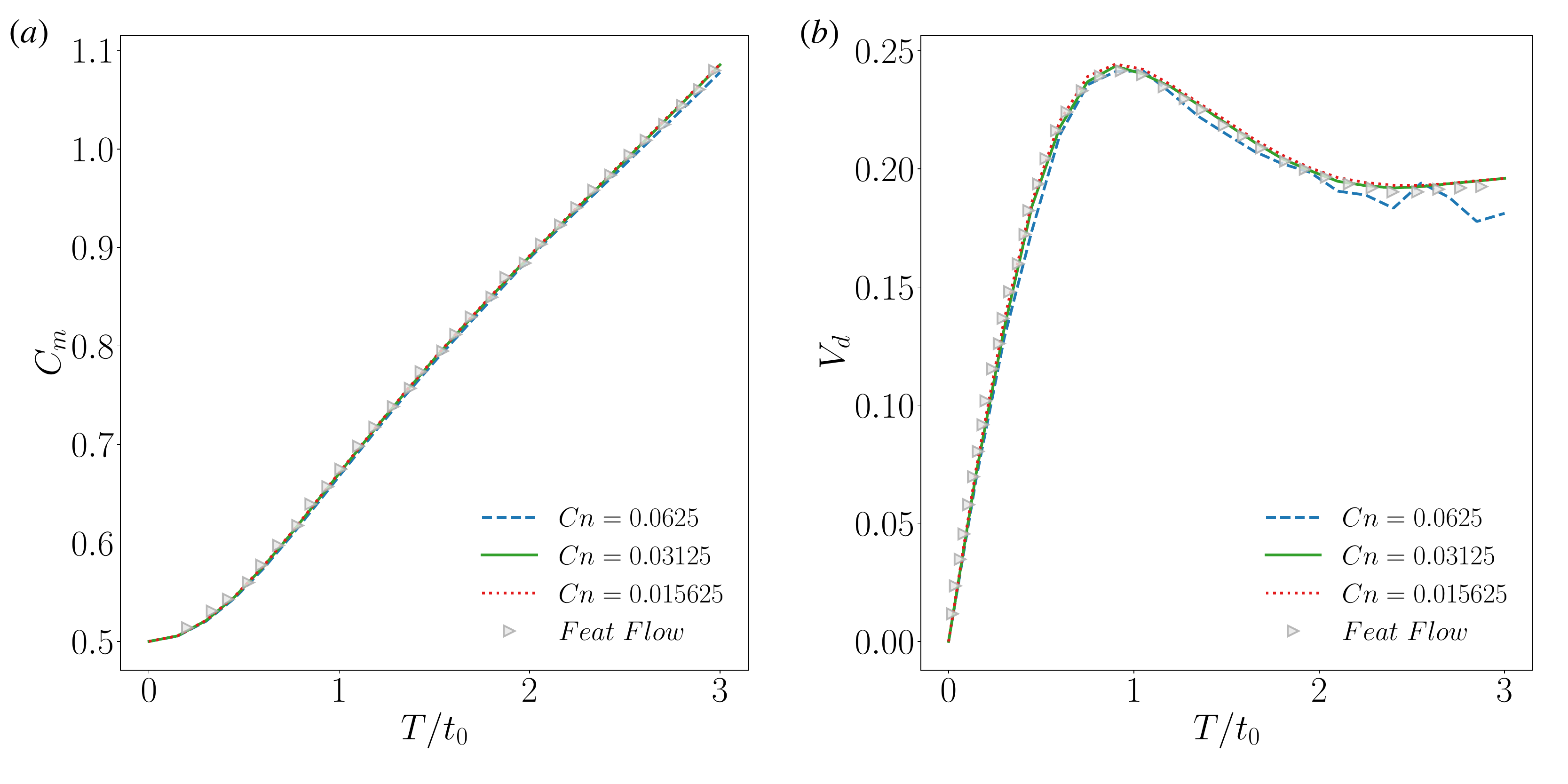}
    \caption{Evolution of (a) the center of mass and (b) vertical velocity when $T/t_0=[0,3.0]$ for $Cn=[0.0625-0.015625]$. The reference curve is provided by $Feat\ Flow$.\ 
    }
\end{figure*}

We apply our model to the validation of two-phase flow simulations with density ratio by considering the single rising bubble benchmark problem. Specifically, we investigate case 1 from Hysing et al. (2009) \cite{hysing2009quantitative}, which is a well-known and widely used benchmark in the field. This benchmark involves the simulation of a single bubble rising in a fluid domain, and it serves as a reliable test to assess the accuracy and performance of two-phase flow models.
\begin{figure}
	\centering
  \includegraphics[width=0.8\linewidth]{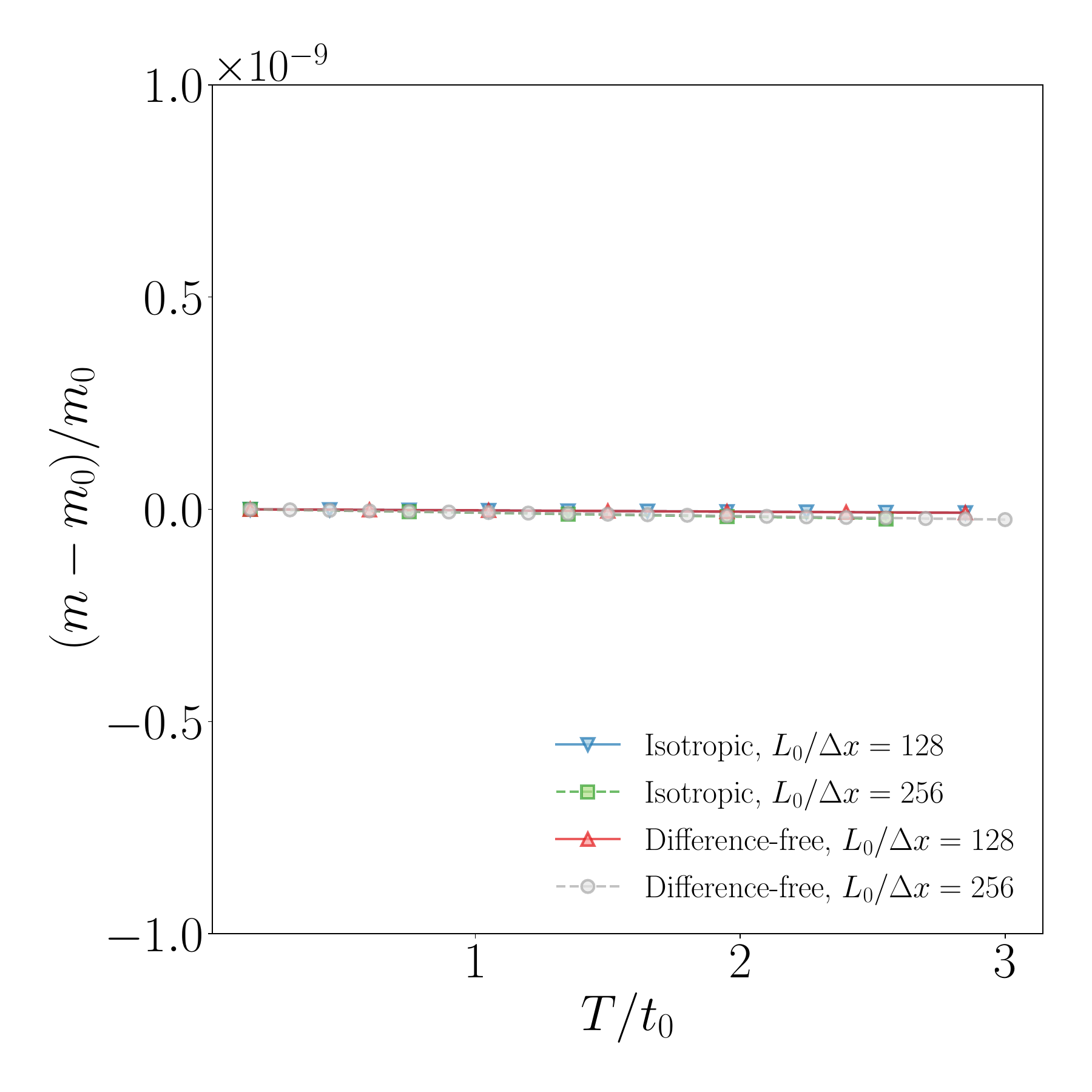}
    \caption{Variation of the mass of enclosed area by the bubble interface versus time.
The lines in green color indicate the results of the CPF lBM model using the isotropic finite difference method and the red symbols show the results based on the current difference-free method. \label{mass}  \ 
    }
\end{figure}
We conduct the single rising bubble simulation in a rectangular pool, where a bubble with lower density is introduced into a higher-density liquid. The dimensions of the rectangular pool are set as $L_0\times2L_0$, and the diameter of the bubble is chosen as $D/L_0=0.5$. The bubble, with a density of $\rho_2$, is initially positioned at $c_m(x,y)/D=(1, 1)$, while the background fluid is filled with liquid of density $\rho_1$. To establish a density contrast, we set the density ratio and viscosity ratio as $\rho_1/\rho_2=\eta_1/\eta_2=10$.

The rising process of the bubble is characterized by the Bond number, which is defined as $Bo=\Delta \rho g D^2/\sigma$, where $\Delta \rho$ is the density difference between the bubble and the liquid, $g$ is the acceleration due to gravity, and $\sigma$ is the surface tension. In our simulation, we consider a Bond number of $Bo=10$. Additionally, we evaluate the Archimedes number, given by $Ar=\sqrt{g D^3}/\nu_1$, where $\rho_1$ and $\nu_1$ represent the density and dynamic viscosity of the liquid, respectively. For this simulation, we set $Ar=35$.

The evolution of the bubble deformation during the rising process is presented in Figure~\ref{rising}(a). To study the effect of the interface thickness, we adjust the Cahn number $Cn$ by varying the length of the rectangular pool while keeping a constant interface thickness $\delta$. By decreasing $Cn$, the morphology of the rising bubble at $T/t_0=3$ converges to the same shape. This indicates that the bubble shape becomes independent of the pool size and is primarily influenced by the interface thickness.

We further analyze the evolution of the mass center $C_m=\sum(\phi y)/\sum \phi$ and the vertical velocity $V_d=\sum(\phi u_y)/\sum\phi$ of the bubble, and compare the results with the reference data from~\cite{hysing2009quantitative}. The center of mass of the simulations with different $Cn$ exhibits highly consistent results. Regarding the evaluation of the vertical velocity, the simulation with lower resolution fails to provide a smooth curve but correctly captures the overall trend. As the resolution increases, the simulations with lower $Cn$ show good agreement with the reference benchmark. The results indicate that by increasing the resolution and reducing the interface thickness, the difference-free scheme produces accurate and reliable simulations, aligning well with the reference data from~\cite{hysing2009quantitative}.

We then compare the mass conservation between the current difference-free method with our previous CPF lBM which employs the isotropic finite difference method~\cite{zhao2023interaction,lee2005stable}. The mass of the droplet is evaluated by $m=\sum\rho\phi\Delta x^3$, and the initial mass is noted as $m_0$. With the evaluation of mass conducted through quadrature, an increase in the interface area between two liquids results in a smaller computed mass. The comparisons of the results at various resolutions $Cn=[0.024,0.047]$ are presented in Figure~\ref{mass}. It is noteworthy that both the isotropic finite difference derivative method and the difference-free method yield very small mass loss. For both methods, the effect of the resolution for mass loss can be neglected.

\subsection{Droplet breakup in a decayed Taylor-Green vortex}

One notable feature of the difference-free numerical scheme is its implicit computation of all derivatives using the local PDF. This characteristic makes the scheme particularly advantageous for large-scale simulations when integrated into distributed computing systems. In order to validate both the efficiency and accuracy of this scheme in such scenarios, we conduct a comparison between the current scheme and our previous method~\cite{zhao2023interaction} using a large-scale simulation of a single droplet breakup inside a decayed Taylor-Green vortex liquid pool.

We initialize a droplet with a diameter of $D/L_0=0.4$ located at $c_m(x,y,z)/L_0=(0.45,0.45,0.45)$ within a three-dimensional Taylor-Green vortex cube. The cube has a constant side length of $L_0$. The initial velocity profile and pressure distribution are defined as follows:

\begin{flalign*}\label{TGV_ini}
\begin{split}
&u(x,y,z,0)  =u_0 \sin\left(\frac{x}{L_0}\right) \cos\left(\frac{y}{L_0}\right) \cos\left(\frac{z}{L_0}\right), \\
&v(x,y,z,0)  =-u_0 \cos\left(\frac{x}{L_0}\right) \sin\left(\frac{y}{L_0}\right) \cos\left(\frac{z}{L_0}\right), \\
&w(x,y,z,0)  =0,\\
&p = p_0+\frac{\rho_0 u_0^2}{16}\left(\cos\left(\frac{2x}{L_0}\right)+ \cos\left(\frac{2y}{L_0}\right)\right)\left(\cos\left(\frac{2z}{L_0}\right)+2\right).
\end{split}
\end{flalign*}
Here, $u_0$ represents the reference velocity, which is determined by the Reynolds number $Re=\rho_0 u_0 L_0 /\eta=1600$. The time scale for this problem is defined as $t_0=L_0/u_0$. Furthermore, a periodic boundary condition is applied to the simulation, ensuring that the flow pattern repeats throughout the computational domain. In this particular test, the surface effect is characterized by the capillary number $Ca=\eta u_0/\sigma$. Convergence tests for velocity and kinetic energy using this method have already been presented in our previous work~\cite{zhao2022ternary}, and thus, we will not present those results again in this context. However, it is important to validate the morphology evolution by examining the calculation of the normal vector, which is solved using Eq. (\ref{norm}). To this end, we compare the results of the morphology evolution under different normal vector calculation methods.

\begin{figure*}
	\centering
  \includegraphics[width=\linewidth]{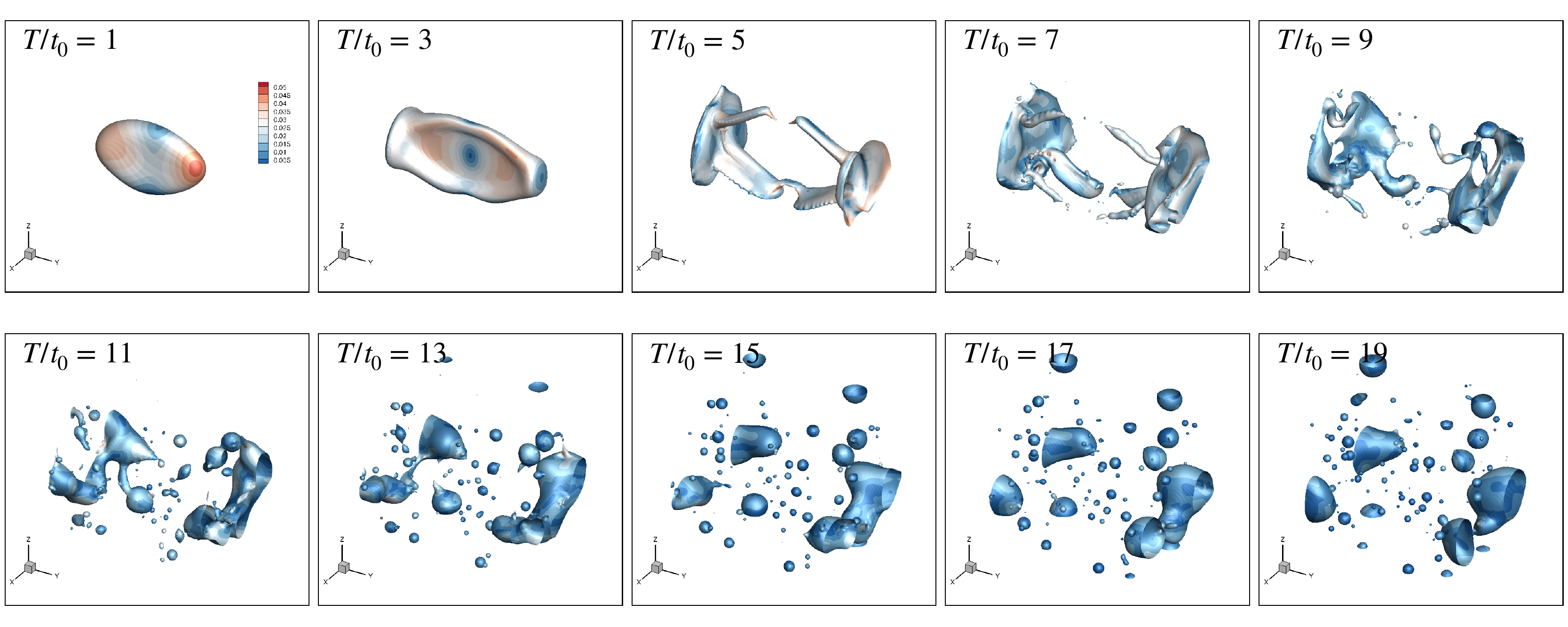}
    \caption{Evolution of the droplet in Taylor-Green vortex by using the difference-free method for $Ca=0.1$ during $T/t_0=[1,19]$. The iso-surface indicates $\phi=0.25$, and the velocity magnitude is shown in contour.  \label{local} 
    }
\end{figure*}

\begin{figure*}
	\centering
  \includegraphics[width=\linewidth]{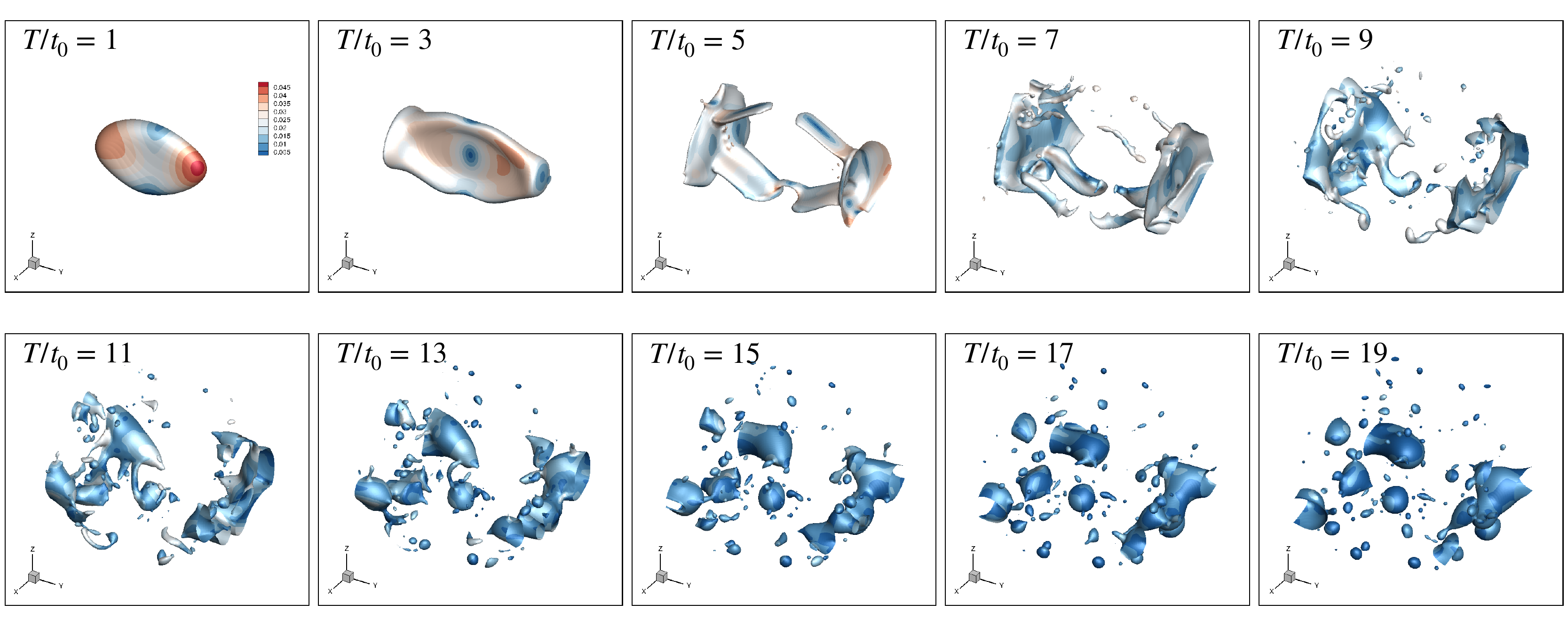}
    \caption{Evolution of the droplet in Taylor-Green vortex by using the isotropic finite difference for $Ca=0.1$ during $T/t_0=[1,19]$. The iso-surface indicates $\phi=0.25$, and the velocity magnitude is shown in contour.  \label{nonlocal} 
    }
\end{figure*}
\begin{figure}
	\centering
  \includegraphics[width=0.8\linewidth]{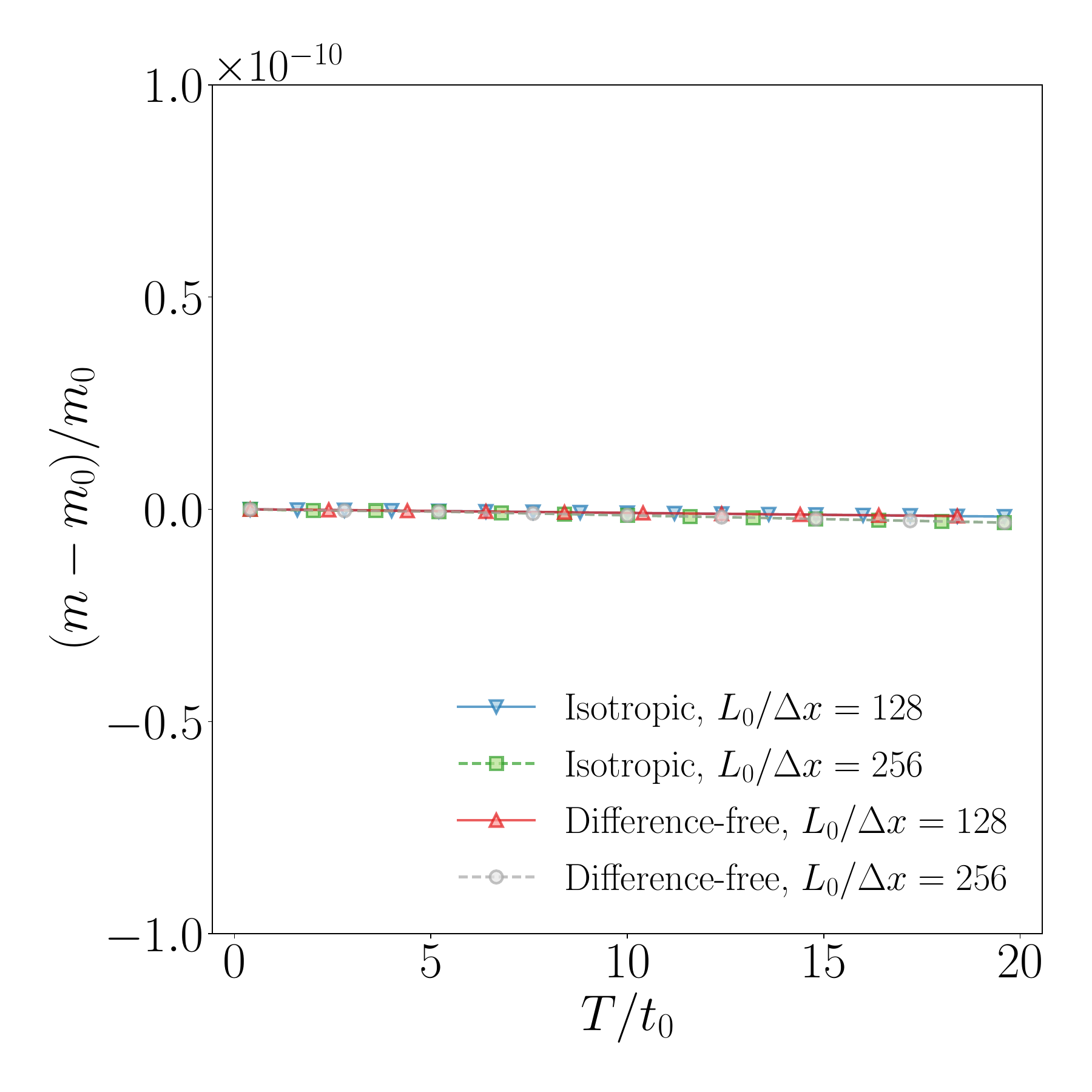}
    \caption{Variation of the mass of the immiscible droplets when $Ca=0.01$.
The lines in green color indicate the results of the CPF lBM model using the isotropic finite difference method and the red symbols show the results based on the current difference-free method. \label{ca001}  \ 
    }
\end{figure}

\begin{figure*}
	\centering
  \includegraphics[width=\linewidth]{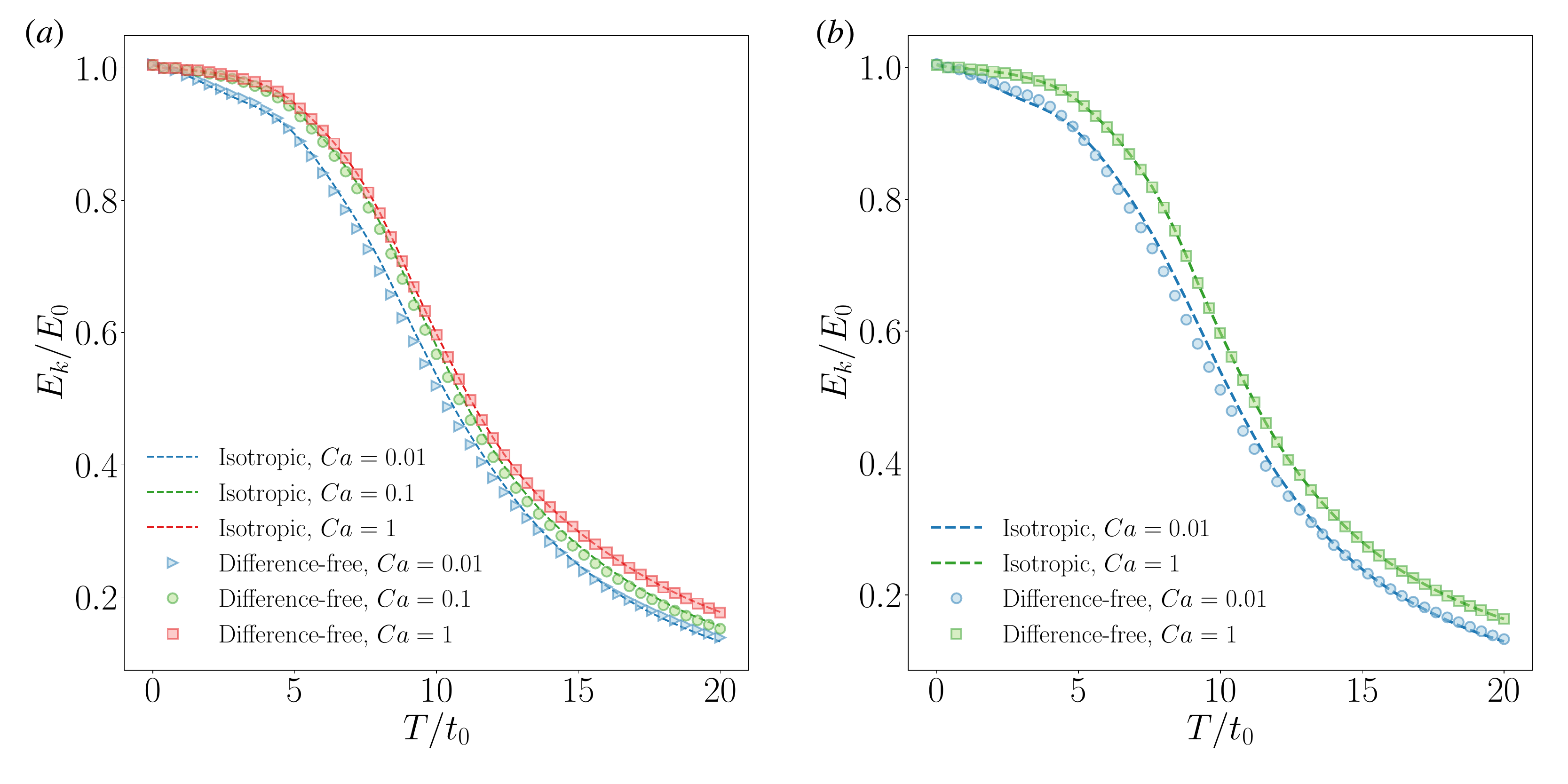}
    \caption{Comparison of the Evolution of the kinetic energy for $Ca=[0.01,1]$ by using the isotropic finite difference derivative method and the difference-free method when $T/t_0=[0,20]$ with (a) $L_0/\Delta x=128$, (b) $L_0/\Delta x=256$. \label{kinetic} 
    }
\end{figure*}
In Figure \ref{local}, we present the results obtained using the difference-free method for computing the normal vector. On the other hand, in Figure \ref{nonlocal}, we utilize the isotropic finite difference method for the normal vector computation, following the expression $\mathbf{n}=\nabla\phi/|\nabla\phi|$ as proposed by Lee et al.~\cite{lee2005stable}.

From the two figures, it can be observed that the initial droplet deforms and eventually breaks up into smaller droplets. At the late stage, due to the surface effect, the droplets evolve into a sphere shape. Overall, the simulation results obtained from both methods show comparability. While it is worth noting that the isotropic finite difference method yields smoother simulation results compared to the difference-free numerical scheme. The presence of some uneven structures in the difference-free scheme can be attributed to the unresolved interface, as discussed in previous studies~\cite{geier2015conservative}. The presence of uneven structures can either advance or delay the breakup of droplets. To mitigate these undesired structures, increasing the Cahn number ($Cn$) and the Courant number ($c$) can be effective measures. The mass conservation is evaluated as the same manner as we did for the rising bubble test case where $m=\sum \rho \phi \Delta x^3$. The relative mass difference is shown in figure~\ref{ca001}, which indicates both methods obtain similar mass loss during the simulation.


We further evaluate the kinetic energy $E_k=\sum\rho\mathbf{u}^2/2$ evolution of both methods. Figure.~\ref{kinetic} shows the comparison of the kinetic energy evolution for both methods with different $Ca=[0.01,1] $ under different resolutions. Without the surface effect, the initial kinetic energy $E_0$ will be fully consumed by the viscosity. As the surface effect is incorporated into the system, a portion of the kinetic energy is consumed by the expanding total surface area, leading to relatively rapid dissipation. According to the comparisons, when $L_0/\Delta x=128$, $Ca=[0.1,1]$, the kinetic energy evolution for both methods is in good agreement. As we increase the resolution to $L_0/\Delta x=256$, both methods obtain similar kinetic evolution for $Ca=1$. When we further increase the surface effect to $Ca=0.01$, the dissipation trend of both methods still remains the same, but a slight deviation is observed during the $T/t_0=[5,10]$. The initial and the final evolution of both cases remain consistent.
\begin{figure*}
	\centering
  \includegraphics[width=0.5\linewidth]{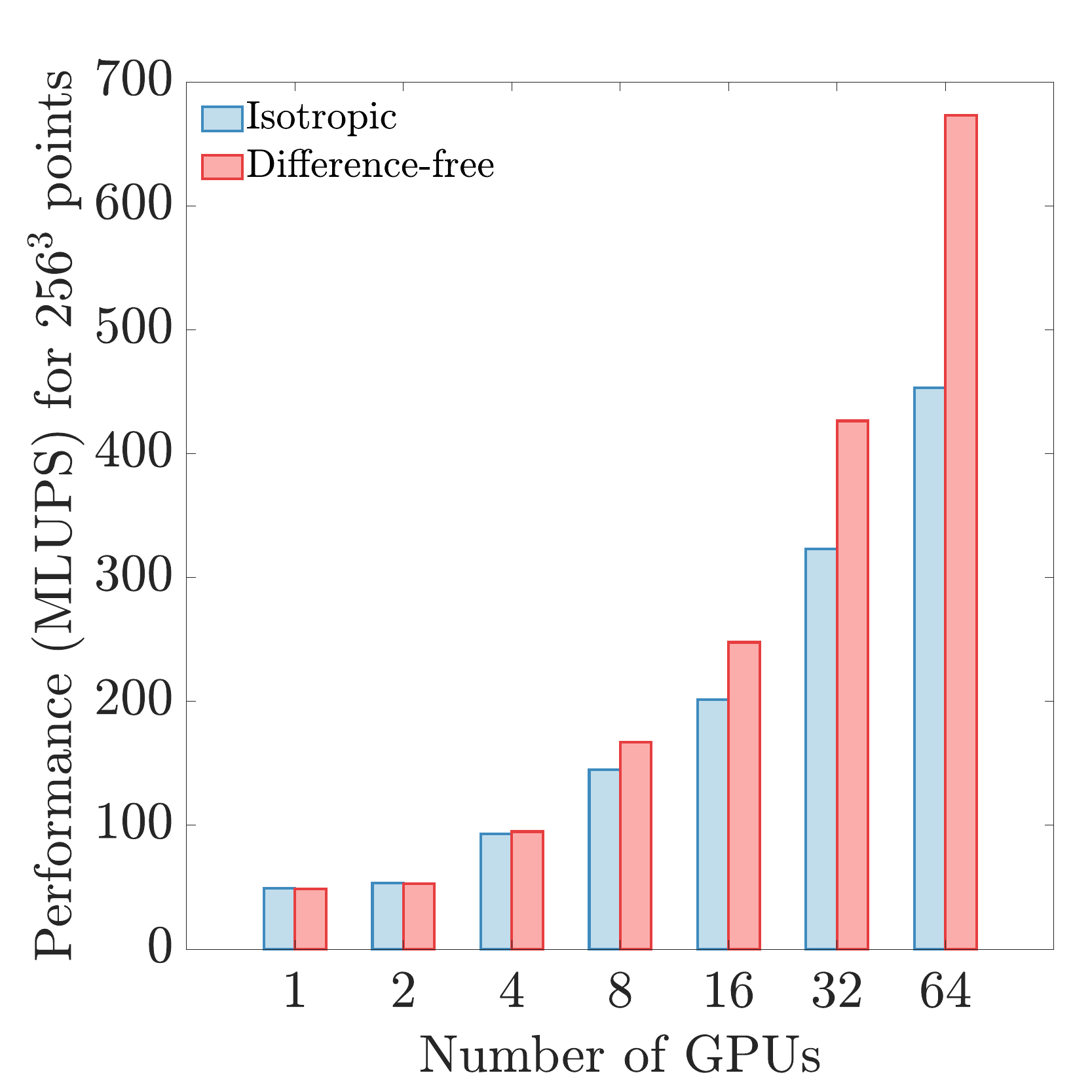}
    \caption{Performance (MLUPS) of isotropic finite difference derivative method and the difference-free method. \label{efficiency} 
    }
\end{figure*}

Finally, We conduct the strong scaling (performance) in terms of Million Floating-Point Operations Per Second (MLUPS) for two schemes, which can be expressed as:
\begin{equation}
    MLUPS=\frac{lattice\ points\ in\ the\ whole\ domain \times iterations }{computing\ time \times 10^6}.
\end{equation}
As shown by this equation, increasing the number of processors reduces the computing time, resulting in improved performance. We conduct the comparison by using IMEXLBM \cite{zhao2022ternary,liu2022imexlbm}
which is open-source software for heterogeneous platforms. In addition, the comparisons are carried out on ThetaGPU \cite{alcfThetagpu}, specifically on the NVIDIA DGX A100. The comparison of the performance is shown in figure~\ref{efficiency}. We maintain a fixed total number of computing grid points at $256^3$ and assess the performance by incrementally increasing the number of GPUs. Consequently, as the number of GPUs is gradually increased, the performance of the difference-free scheme improves progressively compared to the isotropic finite difference method. 
With the utilization of 64 GPUs, a noticeable efficiency increase of $47\%$ is observed.

\section{Concluding remarks}

We have proposed a difference-free multi-phase fluid flow CPF lBM solver. This method offers several advantages, such as the local calculation of density/order parameter derivatives and normal vectors through the central moment of the particle distribution function. Additionally, the surface effect is implicitly incorporated into the equilibrium PDF of the pressure. By combining these numerical schemes, the multi-phase fluid flow can be solved without the need for additional derivative evaluations using finite difference methods.

To validate the proposed method, we conducted various benchmark tests. Firstly, we compared the intensity of spurious currents between the current numerical scheme and different surface force formulations. The results showed that the new method achieved similar outcomes to the previous finite difference based continuous surface stress formulation. Moreover, the relative error of the order parameter demonstrated second-order accuracy, indicating the robustness and accuracy of the method. Furthermore, we investigated classical benchmark problems including Rayleigh-Taylor instability and rising bubble scenarios by increasing the density ratio and viscosity ratio. The simulation results of both problems exhibited high consistency with previous research.

Additionally, we applied the new method to study droplet breakup in a decayed Taylor-Green vortex liquid pool and compared it with our previous approach. The local calculation of derivatives in the new method resulted in improved computational efficiency compared to the previous approach. The droplet distribution and kinetic energy evolution were also in good agreement with the previous method.

However, it is important to acknowledge the limitations of this method, particularly when dealing with sharp interfaces. In cases where the interface thickness is small and cannot adequately resolve the interface region, the method may exhibit some unsmooth structures at the interface. Further research is needed to investigate the mobility and limitations of the method in order to enhance its performance.

Overall, the proposed difference-free multi-phase fluid flow LB solver has shown promising results in various benchmark tests, highlighting its potential for accurate and efficient simulations.

\section*{Acknowledgement}
This research used resources of the Argonne Leadership Computing Facility, which is a DOE Office of Science User Facility supported under Contract DE-AC02-06CH11357. This research was supported by the Exascale Computing Project (17-SC-20-SC), a joint project of the U.S. Department of Energy’s Office of Science and National Nuclear Security Administration, responsible for delivering a capable exascale ecosystem, including software, applications, and hardware technology, to support the nation’s exascale computing imperative. This research was supported by the National Science Foundation under Grant No. 1743794, PIRE: Investigation of Multi-Scale, Multi-Phase Phenomena in Complex Fluids for the Energy Industries.
\clearpage

\appendix
\section{}\label{ap1}
The details of the construction of the one-fluid LBE have been given by Tims Reis~\cite{reis2022lattice}, and we show the simple derivation of this method. We apply the Chapman-Enskog analysis in convection scaling ($\Delta t\sim \Delta x$) from the discrete Boltzmann equation without the body force:
\begin{equation}\label{dbe}
    \frac{\partial f_i}{\partial t}+\mathbf{e}_i\cdot\nabla f_i=-\frac{1}{\lambda_\rho}\left(f_i-f^{eq}_i\right)+S_i.
\end{equation}
The restriction equations can be given as follows:
\begin{equation}
    \sum_i f_i=\sum_i f^{eq}_i=\frac{p}{c_s^2}
\end{equation}

\begin{equation}
    \sum_i f_i \mathbf{e}_i=\sum_i f^{eq}_i \mathbf{e}_i=\rho \mathbf{u}
\end{equation}
\begin{equation}
   \sum_i f_i^{neq}=0, \sum_i f_i^{neq} \mathbf{e}_i=\mathbf{0}
\end{equation}
\begin{equation}
    \sum_i S_i =\mathbf{u}\cdot\nabla\rho.
\end{equation}
In this case, Eq.~(\ref{dbe}) can recover the macroscopic equation:
\begin{equation}
    \frac{1}{\rho c_s^2}\frac{\partial p}{\partial t}+\nabla\cdot \mathbf{u}=0,
\end{equation}
by the zeroth order moment of the PDFs.

We further consider the following expansion series:
\begin{equation}
    f_i=f^{eq}_i+\lambda    _\rho f_i^{1}+ ...
\end{equation}
\begin{equation}
    \mathbf{\Pi}=\mathbf{\Pi}^{eq} +\lambda_\rho\mathbf{\Pi}^{1}+...
\end{equation}
\begin{equation}
    \mathbf{Q}=\mathbf{Q}^{eq} +\lambda_\rho\mathbf{Q}^{1}+...
\end{equation}
\begin{equation}
    \partial t=\partial t_0 +\lambda_\rho \partial t_1+...,
\end{equation}
where tensor $\mathbf{\Pi}=\sum_i f_i \mathbf{e}_i\mathbf{e}_i$ denotes the second order moment of the PDF, and $\mathbf{Q}=\sum_i f_i \mathbf{e}_i\mathbf{e}_i\mathbf{e}_i$ is the third order moment of the PDF.
The DBE equation of the zeroth order of $\tau_\rho$ can be shown as:
\begin{equation}
    \frac{\partial f_i^{eq}}{\partial t_0}+\mathbf{e}_i\cdot\nabla f^{eq}_i=f_i^1,
\end{equation}
and its zeroth order moment is:
\begin{equation}\label{moment0}
   \frac{1}{\rho c_s^2} \frac{\partial p}{\partial t_0}+\nabla \cdot\mathbf{u}=0.
\end{equation}
The first order moment can be computed by multiplying $\mathbf{e}_i$:
\begin{equation}\label{moment1}
    \frac{\partial\rho\mathbf{u}}{\partial t_0}+\nabla\cdot\mathbf{\Pi}^{eq}=0,
\end{equation}
The second moment is then:
\begin{equation}\label{moment2}
    \frac{\partial \mathbf{\Pi}^{eq}}{\partial t_0}+\nabla\cdot \mathbf{Q}^{eq}=-\mathbf{\Pi}^1+\sum_i S_i \mathbf{e}_i\otimes\mathbf{e}_i.
\end{equation}
In order to recover the N-S momentum equation as shown in Eq.~(\ref{ns_m}), the flux term $\mathbf{\Pi}^{eq}$ in Eq.~(\ref{moment1}) needs to be constraint by:
\begin{equation}
    \mathbf{\Pi}^{eq}=\rho\mathbf{u}\otimes\mathbf{u}+p\mathbf{I}-\mathbf{\Pi}_s.
\end{equation}
The construction of the equilibrium PDF can be shown that:
\begin{equation}\label{equilibrium}
    f_i^{eq}=w_i\left(\frac{p}{c_s^2}+\rho\left( \frac{\mathbf{e}_i \cdot \mathbf{u}}{c_s^2}+
\frac{(\mathbf{e}_i\cdot\mathbf{u})^2}{2c_s^4}-
\frac{|\mathbf{u}|^2}{2c_s^2}\right) +\frac{1}{2c_s^4}\mathbf{\Pi}_s \mathbf{:} \left(\mathbf{e}_i\otimes\mathbf{e}_i-c_s^2\mathbf{I}\right)\right).
\end{equation}
With the equilibrium PDF, we can compute the divergence of the third order moment $\nabla\cdot\mathbf{Q}^{eq}$ as:
\begin{flalign*}
    \nabla\cdot\mathbf{Q}^{eq} &=[c_s^2\nabla\cdot \left(\rho  \mathbf{u}\right)]\mathbf{I}+c_s^2\nabla\rho\mathbf{u}+c_s^2\left(\nabla\rho\mathbf{u}\right)^T \\
    &=[c_s^2\nabla\cdot \left(\rho  \mathbf{u}\right)]\mathbf{I}+c_s^2\rho\nabla\mathbf{u}
+c_s^2\mathbf{u}\otimes\nabla\rho
+c_s^2\rho\left(\nabla\mathbf{u}\right)^T
+c_s^2(\mathbf{u})^T\otimes\nabla\rho\\
&=[c_s^2\nabla\cdot(\rho\mathbf{u})]\mathbf{I}+\rho c_s^2\left(\nabla\mathbf{u}+\nabla(\mathbf{u})^T\right)
+c_s^2\mathbf{u}\otimes\nabla\rho+c_s^2\nabla\rho\otimes\mathbf{u}.
\end{flalign*}
The first order temporal derivative of the tensor $\mathbf{\Pi}^{eq}$ can be shown as:
\begin{flalign*}
\frac{\mathbf{\partial \Pi}^{eq}}{\partial t_0}=\frac{\partial \rho \mathbf{u}\otimes\mathbf{u}}{\partial t_0}+\frac{\partial p\mathbf{I}}{\partial t_0}-\frac{\partial\mathbf{\Pi}_s}{\partial t_0},
\end{flalign*}
and:
\begin{align*}
    \frac{\partial \rho \mathbf{u}\otimes \mathbf{u}}{\partial t_0}&=\mathbf{u}\otimes\frac{\partial( \rho\mathbf{u})}{\partial t_0}+\frac{\partial (\rho\mathbf{u}) }{\partial t_0}\otimes \mathbf{u}-
    \frac{\partial \rho}{\partial t_0}\mathbf{u} \otimes \mathbf{u}\\
    &=-\mathbf{u}\otimes\nabla\cdot\mathbf{\Pi}^{eq}-\nabla\cdot\mathbf{\Pi}^{eq}\otimes\mathbf{u}+(\rho c_s^2\nabla\cdot\mathbf{u}) \mathbf{u}\otimes\mathbf{u}\\
    &=-\mathbf{u}\otimes\nabla p-\nabla p \otimes \mathbf{u}+\mathbf{u}\otimes\nabla\cdot\mathbf{\Pi}_s+\nabla\cdot\mathbf{\Pi}_s\otimes \mathbf{u}+\mathcal{O}(Ma^3),
\end{align*}
\begin{equation}
    \frac{\partial p\mathbf{I}}{\partial t_0}=-\rho c_s^2(\nabla\cdot\mathbf{u}) \mathbf{I}.
\end{equation}
With those relations available, we can rearrange Eq~(\ref{moment2}) as:
\begin{align*}
    \mathbf{\Pi}^1&=\rho c_s^2(\nabla\cdot \mathbf{u})\mathbf{I}+\mathbf{u}\otimes\nabla p+\nabla p\otimes\mathbf{u}-\mathbf{u}\otimes \nabla\cdot\mathbf{\Pi}_s-\nabla\cdot\mathbf{\Pi}_s\otimes\mathbf{u}+\frac{\partial\mathbf{\Pi}_s}{\partial t_0 }\\
    &-\rho c_s^2\left(\nabla\mathbf{u}+
    (\nabla\mathbf{u})^T\right)
    +\sum_i S_i\mathbf{e}_i\mathbf{e}_i-
    c_s^2\nabla\cdot(\rho\mathbf{u})\mathbf{I}-c_s^2\mathbf{u}\otimes\nabla\rho-c_s^2\nabla\rho\otimes\mathbf{u}+\mathcal{O}(Ma^3).
\end{align*}
We then construct the forcing $S_i=(\Gamma_i-w_i)(\mathbf{e}_i-\mathbf{u})\cdot\nabla\rho$, the last three terms can be compensated. 
The full second-order tensor can be computed by Eq.~(A8):
\begin{equation}
    \mathbf{\Pi}=\rho\mathbf{u}\otimes\mathbf{u}+p\mathbf{I}-\mathbf{\Pi}_s-\mathbf{\Pi}_v+\mathcal{O}(Ma^3)+\mathcal{O}(Ma^2/Re).
\end{equation}
Finally, Eq.~(\ref{ns_m}) can be recovered.
\section{}\label{ap2}
The details of the derivation can be found in ~\cite{geier2015conservative}. We here present a simple derivation based on the previous approach. To obtain the first derivative of the order parameter, we apply the asymptotic analysis of Eq.~(\ref{dbe_order}) by diffusive scaling ($\Delta t\sim\Delta x^2$). We first denote the central moment of the PDF $g_i$ by:
\begin{equation}
    K_0=\sum_i g_i,    
\end{equation}
\begin{equation}
    \mathbf{K}_1=\sum_i g_i(\mathbf{e}_i-\mathbf{u}).    
\end{equation}
The asymptotic series is then:
\begin{equation}
    K=K^0+\epsilon K^1+\epsilon^2 K^2+...
\end{equation}
where $\epsilon\sim \Delta x$. The Taylor series expansion of the discrete Boltzmann equation Eq.~(\ref{dbe_order}) then can be shown as:
\begin{equation}
    \epsilon^2\frac{\partial g_i}{\partial t}-\epsilon \mathbf{e}_i\nabla\cdot g_i=0,
\end{equation}
 from which we can obtain the moment relation:
\begin{equation}
    \nabla\cdot {\mathbf{\bar{K}}_1^0}=0,
\end{equation}
\begin{equation}
    \frac{\partial K_0^0}{\partial t}=-\nabla\cdot {\mathbf{\bar{K}}_1^1},
\end{equation}
\begin{equation}\label{vK}
 \mathbf{K}_1^1=\mathbf{K}_{1}^{*1}-\nabla\cdot\mathbf{\bar {K}}_2^0,
\end{equation}
Where 
\begin{equation}\label{eqK}
    K^*=K-(K-K^{eq})/\left(\tau_\phi+0.5\right)
\end{equation}
denotes the post-collision state, $\tau_\phi$ is the relaxation time, and $\bar{K}=0.5(K+K^*)$ represents the average value of moments. $\mathbf{K}_2$ takes the form 
\begin{equation}
    \mathbf{K}_2=\sum_i g_i(\mathbf{e}_i-\mathbf{u})(\mathbf{e}_j-\mathbf{u})
\end{equation}
can be shown to be a tensor, while when $i\neq j$, the component of the tensor becomes zero. When we insert Eq.~(\ref{eqK}) into~Eq.~(\ref{vK}), we can obtain:
\begin{equation}
 \mathbf{K}_1^1=\mathbf{K}_1^{eq,1}-(\tau_\phi+0.5)\nabla\cdot\mathbf{\bar{K}}_2^0,
\end{equation}
\begin{equation}
 \mathbf{K}_1^{*1}=\mathbf{K}_1^{eq,1}-(\tau_\phi-0.5)\nabla\cdot\mathbf{\bar{K}}_2^0,
\end{equation}
\begin{equation}
 \mathbf{\bar{K}}_1^{1}=\mathbf{K}_1^{eq,1}-\tau_\phi\nabla\cdot\mathbf{\bar{K}}_2^0.
\end{equation}
When we construct the equilibrium momentum as:
\begin{equation}\label{Keq}
    \mathbf{K}_1^{eq}=\tau_\phi\frac{4\phi(1-\phi)}{\delta}\frac{\nabla\phi}{|\nabla\phi|},
\end{equation}
the moment $\mathbf{K}_1$ is then:
\begin{equation}
    \mathbf{K}_1^1=\Lambda \nabla\phi,
\end{equation}
where $\Lambda$ is a priory parameter. Finally, the normal vector can be approximated as:
\begin{equation}
    \mathbf{n}=-\frac{\mathbf{K}_1}{|\mathbf{K}_1|}.
\end{equation}

From Eq.~(\ref{Keq}), it is also possible to compute the derivative of the order parameter up to second order $\sim\mathcal{O}(\epsilon^2)$:
\begin{equation}
    \nabla\phi=\frac{1}{(\tau_\phi+0.5)c_s^2}\left(\tau_\phi c_s^2\frac{4\phi(1-\phi)}{\delta}\mathbf{n}-\mathbf{K}_1\right).
\end{equation}



 \bibliographystyle{elsarticle-num} 
 \bibliography{cas-refs}

\end{document}